\documentclass[11pt,american]{article}
\usepackage[T1]{fontenc}
\usepackage{babel}
\usepackage{csquotes}
\usepackage{amsmath}
\usepackage{graphicx}
\usepackage{amssymb}
\usepackage[margin=1in]{geometry}
\usepackage{float} 
\usepackage{hyperref}
\usepackage{xcolor}
\usepackage{comment}

\title{Optimal Modulation Current for Gain-Switching Lasers}
\author{Alex Borisevich, \href{mailto:akpc806b@gmail.com}{akpc806b@gmail.com}}

\begin{document}

\maketitle

\section*{Abstract}

This paper formally shows that an exponentially rising current is optimal in terms of resistive ohmic loss for driving a semiconductor laser into the gain-switching mode. A metric to quantify the quality of laser operation that measures the similarity of a generated optical pulse to the delta function is proposed. Several circuit implementations to approximate exponentially rising current are developed, including using a driver circuit with BJT output stage, a network of RLC circuits, and a saturating inductor. An experimental comparison between a state-of-the-art sinewave resonant driver circuit and a directly driven laser is performed that favors the latest variant of the driver.

\section{Introduction}

Gain-switching is a technique used in lasers to generate short-duration optical pulses. In a gain-switched laser, the gain medium (the material that amplifies the light) is modulated or switched on and off rapidly, leading to the emission of pulsed laser light. This is in contrast to continuous-wave (CW) lasers that emit a steady beam of light.

The basic principle behind gain-switching involves rapidly changing the population inversion in the gain medium. Population inversion is a condition where more atoms or molecules in the gain medium are in an excited state than in the ground state, which is necessary for laser amplification to occur. By quickly "switching" the gain medium from a low-population-inversion state to a high-population-inversion state, a burst of photons is emitted as the excited particles rapidly decay back to the ground state, resulting in a short-duration pulse of laser light. By quickly "switching" the gain medium from a low-population-inversion state to a high-population-inversion state, a burst of photons is emitted as the excited particles rapidly decay back to the ground state, resulting in a short-duration pulse of laser light.

Gain-switched lasers are valued for their ability to generate optical pulses with durations ranging from picoseconds to nanoseconds, typically an order of magnitude shorter than applied electrical pulses. These short-duration pulses have applications in telecommunications, laser material processing, spectroscopy, and medical imaging.

It is well known that the electrical-to-optical efficiency of a laser in the gain-switching mode is extremely low, in the order of a couple of percents. In most applications this is acceptable, but in some applications like wearable or portable devices, the efficiency of the laser is critical for battery life and laser parameters stability that drift with a temperature. The laser inefficiency problem is even more amplified in high repetition rates due to linear scaling of the power dissipation with a frequency.

Laser driving circuits for high speed lasers could be loosely divided into two groups: 

1. With a direct electrical coupling of the switching elements to the laser diode, for example \cite{steppulse_modulation}, \cite{direct_drive_prechage}, \cite{direct_drive2}.

2. With a capacitive coupling to the laser where a capacitor is used to store energy which is pumped into the laser, for example \cite{resonant_example}, \cite{resonant_example2}, \cite{resonant_example3}, \cite{resonant_example4}.

There are a number of attempts \cite{steppulse_modulation}, \cite{direct_drive_prechage} to increase efficiency of the laser operation by customizing modulation current profile. The presented approaches are based on staggering modulation current profile into two phases: slow and fast dynamics. The slow part is used to precharge the laser to build up a lasing carrier density threshold, while the fast part which is required to be a very narrow current spike is to inject carriers into precharged lasing laser. 

By our knowledge, the problem of energy efficiency of the gain switching lasers is not systematically addressed in scientific publications and patents, except series of papers \cite{comparision1}, \cite{comparision2}, \cite{comparision3}, \cite{comparision4}. Particularly in \cite{comparision1} it was observed from the numerical simulations that tangential hyperbolic input provides the maximum power along with minimum FWHM.

In this paper we will demonstrate that exponential modulation current is optimal in terms of the electrical losses minimization of the gain switching lasers, along with some concepts how to implement this modulation current by electrical circuits. We also will introduce a metric to evaluate quality of the optical pulses and compare state of the art resonant driver circuit with a circuit that generates more optimal modulation current waveform.

\section{Optimal Precharge Current Profile}

In this section we are going to formally prove that an optimal modulation current to start the lasing in gain-switching mode has a form: $I(t) = A(T,N_{th}) \cdot \exp(t / \tau_N)$ where $A(T,N_{th})$ is a constant that is a function of the pulse duration $T$ and the lasing threshold carrier density $N_{th}$. The exponential growth rate constant $\tau_N$ is a physical characteristic of the laser.

\subsection{State Space Model}

The state-space model \cite{steppulse_modulation} of the gain switching laser is 

\begin{equation}\label{eq:rate}
\begin{gathered}
\dot N = \frac{I}{e V} - \frac{N}{\tau_N} - g(N,S) \\
\dot S =  \Gamma \cdot g(N,S) -  \frac{S}{\tau_P} + \frac{\Gamma \beta N}{\tau_N} 
\end{gathered}
\end{equation}

where $N$ and $S$ are the carrier density and photon density, $I$ is the injected current, all other constants are the laser physical parameters, namely: $\tau_N$ is total spontaneous emission carrier lifetime, $\tau_P$ is average photon lifetime inside the cavity, $\Gamma$ is mode confinement factor, $\beta$ is fraction of the spontaneous emission, $e$ is elementary charge, $V$ is active region volume.

The $g(N,S)$ is a nonlinear term, which models gain of the medium 

\begin{equation}
g(N,S) = g_0\frac{(N - N_t) \cdot S}{1 + \epsilon S}
\end{equation}

where $g_0$ is gain slope constant, $N_t$ is carrier density at transparency, $\epsilon$ is gain compression factor.

\begin{figure}[H]
\centering
\includegraphics[width=0.6\textwidth]{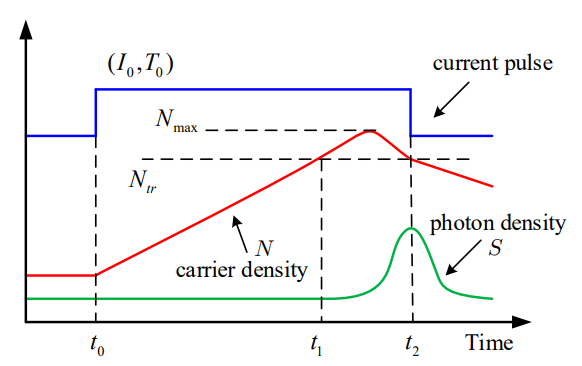}

\caption{Gain switching principle illustration (from \cite{steppulse_modulation})}
\label{fig:Gain_switching_principle}
\end{figure}

The gain switching mode of laser diode operation can be described as follows (Figure \ref{fig:Gain_switching_principle}). A current pulse $I(t)$ is applied to a laser. Without being significantly consumed by stimulated emission, injected electrons rapidly build the carrier density $N$ up to threshold density $N_{th}$. After that, the population inversion can be achieved, and the laser begins to emit the light pulse, which is modeled by the second equation for $S$, and especially first term $\Gamma \cdot g(N,S)$. When the stimulated emission begins to consume the carrier significantly, which is modeled by $-g(N,S)$ term in the first equation, the population inversion reaches its maximum value. With the generation of laser pulses, the carrier density $N$ drops to the lasing threshold, at which point the emission $S$ reaches its peak value. At this time, the current should be terminated quickly to restrain the secondary optical oscillation.

The lasing threshold carrier density is defined by a condition, when the medium gain is higher than cavity losses:

\begin{equation}
N_{th} = N_t + \frac{1}{\tau_P \Gamma g_0}
\end{equation}

The minimum current to achieve the laser effect in steady state is called the threshold current $I_{th}$:

\begin{equation}\label{eq:I_th}
I_{th} = \frac{e V}{\tau_N} N_{th}
\end{equation}

\subsection{Deriving the Optimal Trajectory}

The dynamics of the carrier density before reaching the lasing threshold can be approximated linearly by letting $g(N,S) = 0$ which gives:

\begin{equation}\label{eq:rate_linear}
\dot N \approx \frac{I}{e V} - \frac{N}{\tau_N}
\end{equation}

An approximate optimal control problem can be formulated as follows:

- find a current trajectory $I^*(t)$ which reaches the threshold carrier density $N_{th}$ by an arbitrary time $T$,

- minimize dissipated in the laser diode energy by minimizing the following performance function:

\begin{equation}\label{eq:perf_I}
J = \int_0^{T} I^2(t) dt \to \min
\end{equation}

And the linear approximation of carrier density $N$ dynamics is a just first order linear dynamical system:

\begin{equation}\label{eq:plant_N}
\dot x = - a x + u / b
\end{equation}

where $x = N$ is state variable, $u = I$ is control input, $a,b > 0$ are coefficients:

\begin{equation}
\begin{gathered}
a = \frac{1}{\tau_N} \\
b = e V
\end{gathered}
\end{equation}

The performance function becomes:

\begin{equation}
J = \int_0^{T} u^2(t) dt \to \min
\end{equation}

Substitution of $u = b \dot x + a b x$ to $J$ gives:

\begin{equation}
J = \int_0^{T} (b \dot x + a b x)^2 dt \to \min
\end{equation}

the corresponding cost functional (integrand) is: 

\begin{equation}
V = (b \dot x + a b x)^2 = a^2 b^2 x^2 + 2 a b^2 \dot x x + b^2 (\dot x)^2 
\end{equation}

In order to find the optimal trajectory for $x$, the Euler-Lagrange equation needs to be satisfied \cite{optimal_control}:

\begin{equation}\label{eq:euler_lagrange}
\frac{\partial V}{\partial x} - \frac{d}{dt} \frac{\partial V}{\partial \dot x} = 0
\end{equation}

over the optimal trajectory $x^*(t)$.

Substituting \eqref{eq:plant_N} into \eqref{eq:euler_lagrange} and simplifying, the corresponding Euler-Lagrange equation becomes a second-order linear ordinary differential equation:

\begin{equation}
\ddot x - a^2 \cdot x = 0 
\end{equation}

The solution of which equations is 

\begin{equation}
x(t) = C_1 e^{a t} + C_2 e^{-a t}
\end{equation}

where the constants $C_1$ and $C_2$ evaluated using the given boundary conditions: $x(0) = 0$, $x(T) = N_{th}$.

\begin{equation}
x(t) = x(T) \frac{\sinh(a t)}{\sinh(a T)}
\end{equation}

The optimal control input (current) trajectory can be found as:

\begin{equation}
u = b \cdot ( \dot x + a x )
\end{equation}

which gives:

\begin{equation}
u(t) = \frac{b \cdot x(T)}{\sinh(a T)} \cdot \left ( a \cosh(a t) + a \sinh(a t) \right )
\end{equation}

and after obvious simplifications:

\begin{equation}
u(t) =  \frac{a \cdot b \cdot x(T)}{\sinh(a \cdot T)} e^{a t}
\end{equation}

or in original quantities:

\begin{equation}\label{eq:optim_current}
I(t) = \frac{e V N_{th}}{\tau_N \sinh(T / \tau_N)} \exp (t / \tau_N)
\end{equation}

\begin{figure}[H]
\centering
\includegraphics[width=0.8\textwidth]{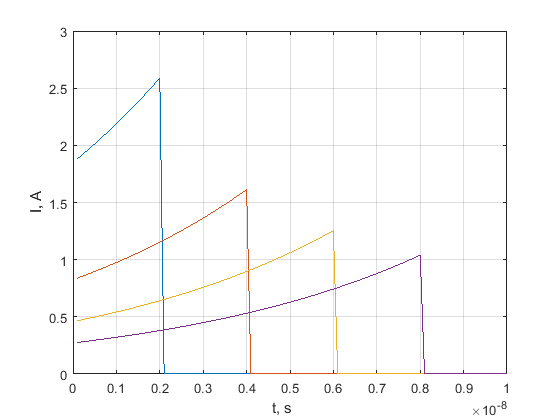}

\caption{Examples of optimal current profiles \eqref{eq:optim_current} for various pulse durations $T$}
\label{fig:optimal_current_profiles}
\end{figure}

Examples of optimal current profiles \eqref{eq:optim_current} for various pulse widths $T$ are demonstrated in Figure \ref{fig:optimal_current_profiles} for a model of 0.5 W semiconductor laser diode.

\subsection{Pulse Duration}

An arbitrary (but fixed) duration $T$ of the control pulse is used in the optimal control problem setting \eqref{eq:perf_I}. It is interesting to find out what duration $T$ is optimal for the energy efficiency.

The performance function $J$ can be evaluated by substituting optimal control $u(t)$ given by equation \eqref{eq:optim_current}:

\begin{equation}
J(T) = \int_0^{T} u^2(t) dt = \frac{a^2 b^2 x(T)^2}{\sinh^2(a T)} \int_0^{T} e^{2 a t} dt = a b^2 x(T)^2\frac{e^{a T}}{\sinh(a T)} =  \frac{e^2 V^2 N_{th}^2 \cdot \exp (T / \tau_N)}{\tau_N  \sinh(T/\tau_N)}
\end{equation}

\begin{figure}[H]
\centering
\includegraphics[width=0.8\textwidth]{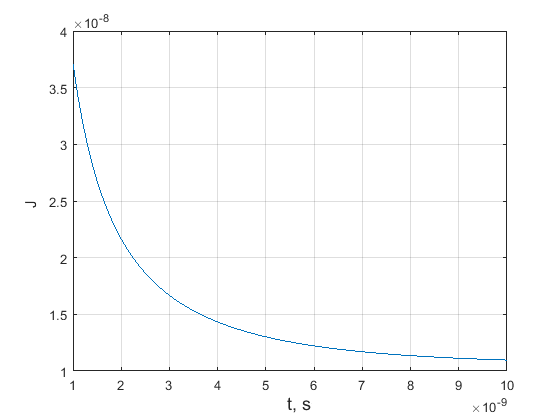}

\caption{Energy loss performance \eqref{eq:perf_I} as a function of pulse duration $T$}
\label{fig:loss_function_of_T}
\end{figure}

The obtained function $J(T)$ can be evaluated for different $T$, and the results are given in the Figure \ref{fig:loss_function_of_T}.

From the $J(T)$ dependency follows that longer pulses are more optimal from the minimizing electrical losses point of view. It can be formally proven that there is a lower limit of the energy loss obtained for an infinite long pulse:

\begin{equation}
J_{\min} = \lim_{T \to \infty} J(T) = \frac{e^2 V^2 N_{th}^2}{\tau_N} \cdot \lim_{T \to \infty} \frac{\exp (T / \tau_N)}{\sinh(T / \tau_N)} = 2 \frac{e^2 V^2 N_{th}^2}{\tau_N}
\end{equation}

However, the optical power is proportional to the instantaneous current in the moment of lasing generation. Thus, there is a trade-off between the optimality of conduction loss and optical power: shorter pulses give higher optical output, but are less efficient.

This can be investigated by varying $T$ and simulating the full nonlinear model \eqref{eq:rate}, calculating efficiency as a quantity proportional to

\begin{equation}\label{eq:eff_sim}
\eta \sim \dfrac{\displaystyle \int_0^{\infty} S(t) dt}{\displaystyle \int_0^{T} I^2(t) dt}
\end{equation}

The results of evaluating the optimal trajectories using the full laser model \eqref{eq:rate} are presented in the figures \ref{fig:current_trajectories_full_model} and \ref{fig:efficiency_full_model} below.

\begin{figure}[H]
\centering
\includegraphics[width=1.1\textwidth]{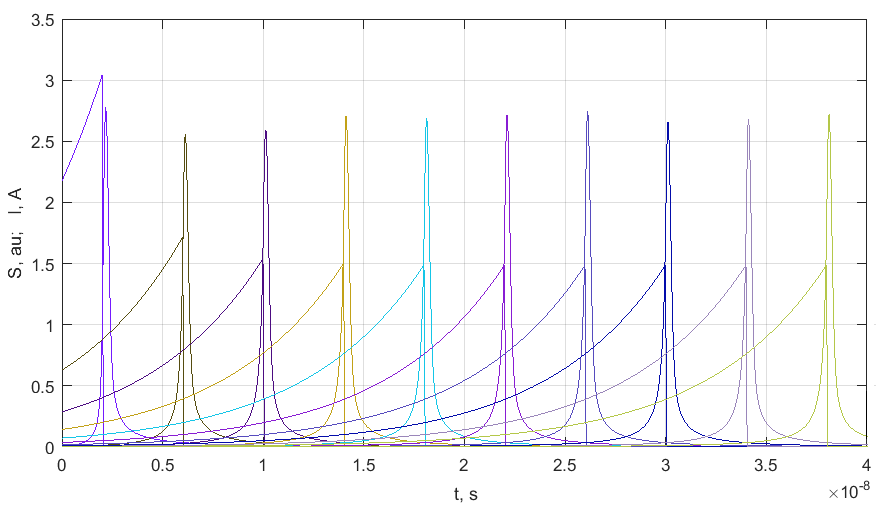}

\caption{Optimal current trajectories \eqref{eq:optim_current} and optical pulses obtained by integrating \eqref{eq:rate} for various pulse durations $T$}
\label{fig:current_trajectories_full_model}
\end{figure}

\begin{figure}[H]
\centering
\includegraphics[width=0.8\textwidth]{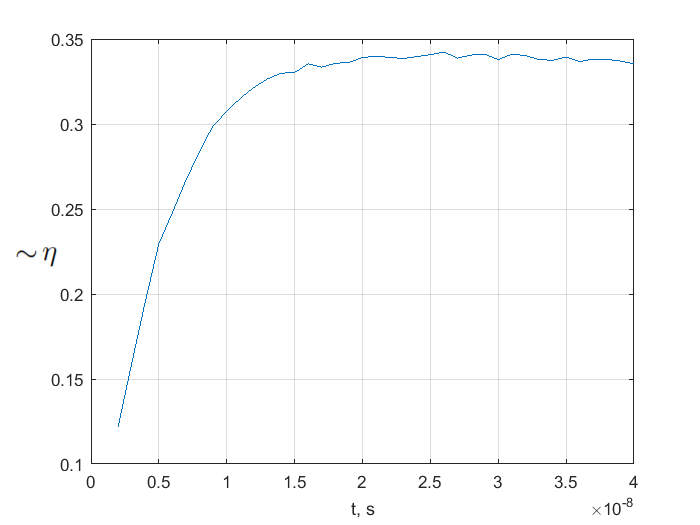}

\caption{Efficiency measure $\eta$ as a function of pulse duration $T$ calculated by integrating \eqref{eq:rate}}
\label{fig:efficiency_full_model}
\end{figure}

It is noticeable from the Figure \ref{fig:current_trajectories_full_model} that the peak current of the optimal control becomes constant starting from some moment of time $T$.
This fact can be formally established by considering the limit

\begin{equation}
\lim_{T \to \infty} I(T) = \frac{e V N_{th}}{\tau_N} \lim_{T \to \infty}\frac{\exp (T / \tau_N)}{\sinh(T / \tau_N)} = 2 \frac{e V N_{th}}{\tau_N} = 2 I_{th}
\end{equation}

Thus, by increasing $T$ the peak current approaches a constant value equal to twice the threshold generation current \eqref{eq:I_th}.

There are two conclusions from the analysis of energy losses as a function of pulse duration $T$:

1. The optimal peak current $I(T)$, starting from sufficiently long pulses approaches $2 I_{th}$.

2. The efficiency ratio $\eta$ as \eqref{eq:eff_sim} increases with increasing pulse duration and also becomes constant, starting from sufficiently long current pulses.

In addition, the physical feasibility of the optimal current waveform implies a limitation on the rate of change of current due to the presence of parasitic inductances in the circuit:

\begin{equation}\label{eq:slew_limit}
\left | \frac{d}{dt} I(t) \right | \le (dI/dt)_{max}
\end{equation}

To meet the slew rate limitation requirement \eqref{eq:slew_limit} note that the time derivative of the exponential function only increases with time $\dot I(T) \ge \dot I(t)$, so it is enough to check only $\dot I(T)$

\begin{equation}
\dot I(T) = \frac{e V N_{th}}{\tau_N^2 \sinh(T / \tau_N)} \exp (T / \tau_N) \le (dI/dt)_{max}
\end{equation}

Solving the inequality above, the duration $T$ should be longer than

\begin{equation}
T \ge \tau_N \sqrt{\frac{B}{B - 2}}
\end{equation}

where

\begin{equation}
B = \frac{\tau_N^2}{e V N_{th}} \cdot (dI/dt)_{max}
\end{equation}

Based on the numerical and analytical fact that the longer the current pulse, the more optimal it is in terms of loss energy $J$, it becomes important to minimize parasitic inductances in the circuit to turn off the current quickly after time $T$ to avoid optical afterpulsing.

\section{An Optical Pulse Shape Metric}

In order to evaluate shape of produced optical pulses a metric can be proposed which is the amplitude (peak) of the signal normalized by the pulse energy. In other words, for a given signal $f(t)$ of finite duration and finite energy, the metric is defined as a ratio of a maximum value of (the $\infty$-norm) to the integral (the 1-norm) of the signal:

\begin{equation}\label{eq:rho}
\rho[f] = \frac{\|f\|_{\infty}}{\|f\|_{1}} = \dfrac{\displaystyle \max_{t \in [0,T]} f(t)}{\displaystyle \int_0^T f(t) dt}
\end{equation}

The definition of the metric is inspired by a crest factor or peak-to-RMS ratio \cite{crest_factor}, which is used for signal waveforms characterization in electrical engineering

For a uniformly sampled discrete signal $y_k$, the integral is obviously just a sum of the signal samples over its duration, so the discrete time version of the $\rho$ is

\begin{equation}
\rho[y] = \dfrac{\displaystyle \max_{k} y_k}{\Delta T \sum_k y_k}
\end{equation}

where $\Delta T$ is sampling interval.

As it seen, the unit of $\rho$ is Hz or 1/s.

\subsection{Mathematical Properties of $\rho$}

1. It is obvious that the $\rho$ metric value calculated for delta function is infinite:

\begin{equation}
\rho[\delta(t)] = \infty
\end{equation}

This is pretty apparent from the definition of the delta function $\delta$:

\begin{equation}
\rho[\delta(t)] = \dfrac{\max_t \delta(t)}{\int_{-\infty}^{+\infty} \delta(t) dt} = \max_t \delta(t) = \delta(0) = \infty
\end{equation}

So the higher the $\rho$ the closer the signal to a delta function shape.

2. The value of $\rho$ metric calculated for a rectangular shape pulse $1(t)$ of unity amplitude and finite duration $T$ equals to $1/T$:

\begin{equation}
\rho[1(t)] = \dfrac{\max_t 1(t)}{\int_{0}^{T} 1(t) dt} = \frac{1}{T}
\end{equation}

3. For two signals $f(t)$ and $g(t)$ of the same energy and finite duration the value of metric $\rho$ will be higher for the signal of higher amplitude.

Let $\int_{0}^{T} f(t) = \int_{0}^{T} g(t) = E$, and also $f_{max} = \max_t f(t)$, $g_{max} = \max_t g(t)$. Assume $f_{max} > g_{max}$, then

\begin{equation}
\rho[f] = \dfrac{\max_t f(t)}{\int_{0}^{T} f(t)} = \dfrac{f_{max}}{E} > \dfrac{g_{max}}{E} = \dfrac{\max_t g(t)}{\int_{0}^{T} g(t)} = \rho[g]
\end{equation}

4. For two signals $f(t)$ and $g(t)$ of the same amplitude and finite duration the value of metric $\rho$ will be higher for the signal with lower energy (which usually corresponds to a shorter in duration pulse).

Let $\max_t f(t) = \max_t g(t) = A$, and also $\int_{0}^{T} f(t) = E_{f}$, $\int_{0}^{T} g(t) = E_{g}$. Assume $E_{f} > E_{g}$, then

\begin{equation}
\rho[f] = \dfrac{\max_t f(t)}{\int_{0}^{T} f(t)} = \dfrac{A}{E_{f}} < \dfrac{A}{E_{g}} = \dfrac{\max_t g(t)}{\int_{0}^{T} g(t)} = \rho[g]
\end{equation}

The above properties indicate that a narrower pulse with a higher peak amplitude will have a higher $\rho$ metric value.

\subsection{Properties of $\rho$ With Respect to System Responses}

Without loss of generality, all signals under consideration are assumed non-negative: $f(t) \ge 0$.

1. The metric $\rho$ is invariant for an arbitrary amplitude scaling of the signal:

\begin{equation}
\rho[k \cdot f] = \rho[f]
\end{equation}

where $k$ is a scalar constant.

It is apparent from \eqref{eq:rho} that:

\begin{equation}
\rho[k \cdot f] = \dfrac{\max_{t} k f(t)}{\int_0^T k f(t) dt} = \dfrac{k \cdot \max_{t} \cdot f(t)}{k \cdot \int_0^T \cdot f(t) dt} = \dfrac{\max_{t} \cdot f(t)}{\int_0^T \cdot f(t) dt} = \rho[f]
\end{equation}

2. The value of metric $\rho$ calculated for the output response of a linear time-invariant system does not exceed the value of metric $\rho$ calculated for the system input signal.

Let's $h$ is an impulse response of a linear time-invariant system. In other words to output response of the system to an input signal $f(t)$ is a convolution $(h*f)(t)$.

Theorem 1.

\begin{equation}
\rho[h * f] \le \rho[f]
\end{equation}

Proof.

Proposition 1. $\|h * f\|_{\infty} \leq \|f\|_{\infty} \cdot \|h\|_{1}$.

This proposition is a particular case of classical Young's convolution inequality: $\|f*g\|_{r} \leq \|f\|_{p} \cdot \|g\|_{q}$, $\frac {1}{p} + \frac {1}{q}=\frac {1}{r}+1$ obtained for $p = r = \infty$ and $q = 1$, and redefining $g := h$.

Proposition 2. $\|h * f\|_{1} = \|f\|_{1} \cdot \|h\|_{1}$

By defenition of convolution

$$\|h * f\|_1 = \int \left| \int f(\tau) h(t-\tau) d\tau \right| dt$$

where the time integrals are calculated for their corresponding time intervals.

The absolute value can be propagated under the integral and across the terms of multiplications assuming $f,g \ge 0$:

$$\int \left| \int f(\tau) h(t-\tau) d\tau \right| dt = \int \int |f(\tau)| \cdot |h(t-\tau)| d\tau dt $$

Using Fubini's theorem, the order of double integration can be changed, which formally gives a product of norms:

$$\int \int |f(\tau)| \cdot |h(t-\tau)| d\tau dt = \int |f(\tau)| d\tau \cdot \int |g(t-\tau)| dt = \|f\|_{1} \cdot \|h\|_{1}$$

Then the theorem can be trivially proven using the propositions 1 and 2:

\begin{equation}
\rho[f*h] = \frac{\|f*h\|_{\infty}}{\|f*h\|_{1}} = \frac{\|f*h\|_{\infty}}{\|f\|_{1} \cdot \|h\|_{1}} \le \frac{\|f\|_{\infty} \cdot \|h\|_{1}}{\|f\|_{1} \cdot \|h\|_{1}} =  \frac{\|f\|_{\infty}}{\|f\|_{1}}  = \rho[f]
\end{equation}

3. If a signal $f_{a^*}$ from a set of signals $F = \{f_a\}$ is optimal in terms of metric $\rho$, then it remains optimal among all the responses of a linear time-invariant system $h$ to all the signals $f_a \in F$. I.e. a linear time-invariant system preserves optimality of the signals in terms of $\rho$.

Theorem 2.

Let's consider a set of signals $F = \{f_a\}$ with a signal $f_{a^*}$ optimal in terms of $\rho$. When passing all the signals in the $F$ through a system $h$, the response $h * f_{a^*}$ to the signal $f_{a^*}$ will be optimal too:

\begin{equation}
\rho[f_{a^*}] \ge \rho[f_a] \Longrightarrow \rho[h * f_{a^*}] \ge \rho[h * f_a]
\end{equation}

The proof is based on the following construction.
Every non-optimal signal except the $f_{a^*}$ in the set $F = \{f_a\}$ can be considered as a response of a specially constructed LTI system $h_a$ to the optimal signal $f_{a^*}$. In other words, the $h_a$ can be found as a deconvolution solution of the equation:

\begin{equation}
f_a = h_a * f_{a^*}
\end{equation}

Then the statement of the theorem can be formulated as follows:

\begin{equation}
\rho[h * f_{a^*}] \ge \rho[h * h_a * f_{a^*}] = \rho[h * f_a]
\end{equation}

Substituting $g := h * f_{a^*}$ and using commutatively of the convolution operation, the following can be obtained for the set of $h_a$ systems:

\begin{equation}
\rho[g] \ge \rho[h_a * g]
\end{equation}

which is a true statement according to the previously proved theorem 1.

A practical outcome of the deduced properties is that the metric $\rho$ can be experimentally measured using limited bandwidth equipment and passing the laser light through a bulk medium. If the parameters of the laser are found optimal using non-ideal equipment, then the same parameters are optimal when using ideal equipment. And the amplitude of the laser optical pulse can be arbitrary scaled by the measurement equipment and method (assuming sufficient SNR or course).

\section{Hardware Circuit Implementations}

The following is the review of possible practical implementations of the high speed current generators with an output current waveform that approximates the optimal one \eqref{eq:optim_current}.

\begin{figure}[H]
\centering
\includegraphics[width=0.3\textwidth]{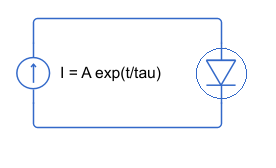}

\caption{A concept of the optimal laser driver as an exponential current source}
\label{fig:laser_driver_concept}
\end{figure}

\subsection{Implementation 1: Push-Pull BJT Driver}

The emitter current in active mode of BJT transistor is modeled by an approximation to the Ebers–Moll model:

\begin{equation}
I_E = I_{ES} \left ( e^{ \frac{V_{BE}}{V_T} } - 1 \right )
\end{equation}

where $V_T = kT/q$ is the thermal voltage (approximately 26 mV at 300 K), $I_E$ is the emitter current, $V_{BE}$ is the base–emitter voltage, $I_{ES}$ is the reverse saturation current of the base–emitter diode, which is a device parameter.

Using this fact, it is straightforward to develop an exponential current generation circuit by applying linearly increasing base–emitter voltage $V_{BE}$.

\begin{figure}[H]
\centering
\includegraphics[width=0.8\textwidth]{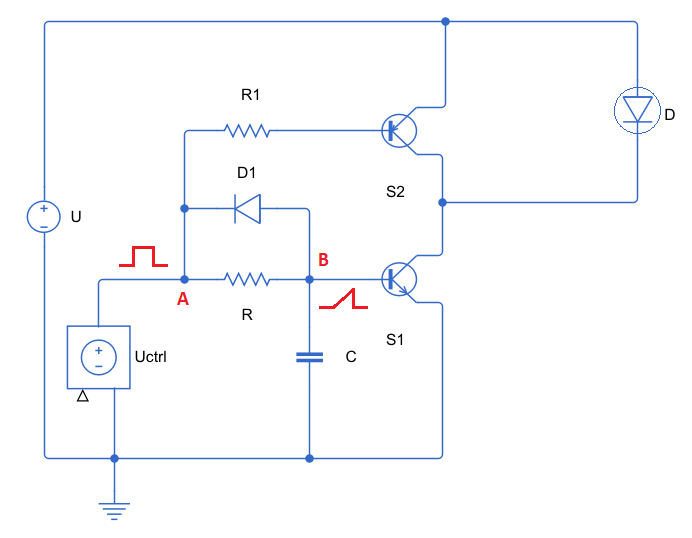}

\caption{Driver topology with BJT transistor as an exponential current source}
\label{fig:BJT_transistor_driver}
\end{figure}

The circuit presented in Figure \ref{fig:BJT_transistor_driver} consists of two complementary BJT transistors S1 and S2. 
During the turn-on phase, the current flows through the laser diode D and turned on transistor S1.
During the turn-off phase, the laser diode D is being shorted through the transistor S2.
The Uctrl is a voltage source of rectangular pulses ranging from 0 V to the power supply voltage U.

The circuit operates as follows: when the voltage at point A becomes equal to the power supply voltage U, the voltage at point B starts rising because of RC circuit consisting of circuit elements R and C. Approximately, the variation of the voltage at the node B is linear (at short time scale). According to the Ebers–Moll equation, the current through the laser diode starts rising exponentially.
In the turn-off phase which is timed to turn off the laser diode just after first optical pulse, the voltage at point A is being driven to 0 V and the transistor S2 is switched on and the laser diode current is shorted through it. At the same time, the base of transistor S1 is being driven through the diode D1 to a 0 V potential, and thus the transistor S1 is being switched off. The diode D1 is select such that the forward voltage drop of it is lower than the base-emitter threshold voltage of S1.
The resistor R1 is used to balance delays of turning on the transistor S2 and switching off the transistor S1 to minimize the shoot-through current.

\begin{figure}[H]
\centering
\includegraphics[width=0.8\textwidth]{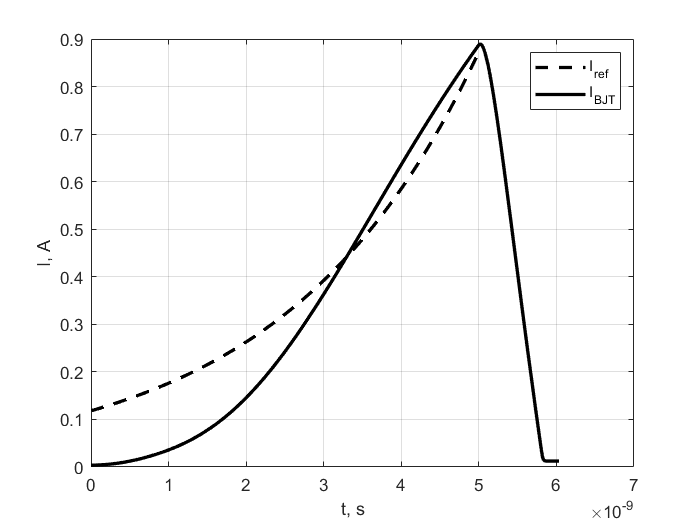}

\caption{Simulated current response of the proposed topology ($I_{BJT}$) in comparison to the optimal trajectory $I_{ref}$ calculated for a 1W 850nm diode model for pulse duration of 5 ns}
\label{fig:BJT_simulated current}
\end{figure}

\subsection{Implementation 2: Multi-Resonant Network}

The idea behind the multi-resonant approach is to superpose multiple sine wave currents. If multiple resonant circuits are loaded to the common resistive load, the current can be approximated as

\begin{equation}
I(t) = -V_0 \sum_i A_i \sin(\omega_i t) = -V_0 \sum_i \dfrac{\sin(t / \sqrt{L_i C_i})} { \sqrt{L_i / C_i} }
\end{equation}

assuming neglecting damping factor and sufficiently small resistive load, and where $V_0$ is an initial voltage of the capacitors.

By a proper selection of inductors $L_i$ and capacitors $C_i$, the current $I(t)$ can be made sufficiently close to the optimal one \eqref{eq:optim_current}.

\begin{figure}[H]
\centering
\includegraphics[width=0.8\textwidth]{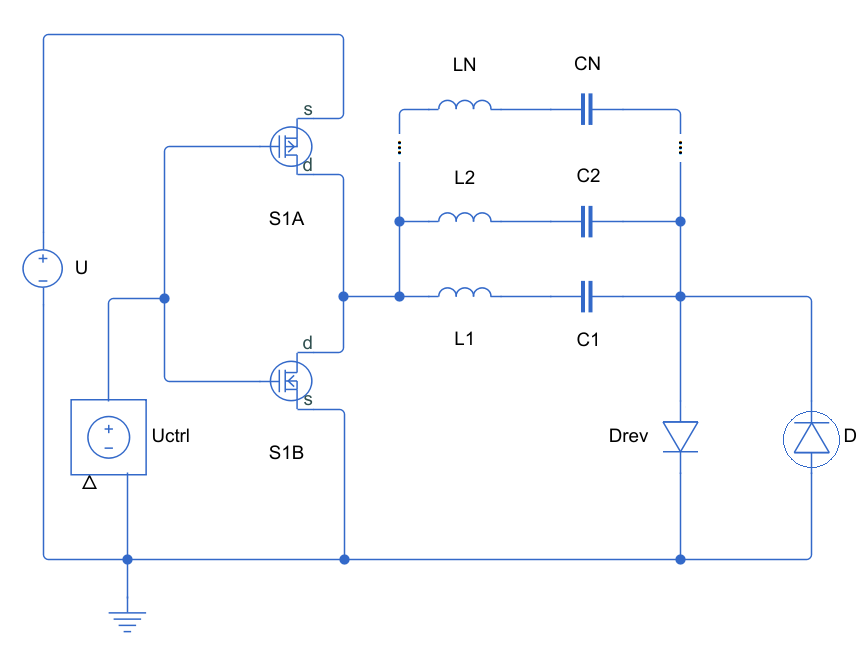}

\caption{Practical circuit of the laser driver with multi-resonant LC circuit}
\label{fig:multi_resonant_LC_circuit}
\end{figure}

A practical circuit of the laser driver which uses this principle is shown in Figure \ref{fig:multi_resonant_LC_circuit}.

The circuit consists of push-pull transistor stage S1A and S1B. The output of the push-pull state is connected to a battery of multiple LC circuits L1,C1, L2,C2 and so on, all connected in parallel. The laser diode D is connected reversely, anode to ground. A discrete diode Drev is connected anti-parallel to the laser diode to charge the capacitors in LC circuits.

The circuit operates as follows: S1A is on initially and the capacitors C1, C2,... CN are charging through the Drev.
Once the capacitors are charged, then the S1A is switched off and the S1B is switched on. Since the S1B is bidirectionally conductive, the current is flowing from its drain to source through the laser diode D and the battery of LC circuits.

\begin{figure}[H]
\centering
\includegraphics[width=0.8\textwidth]{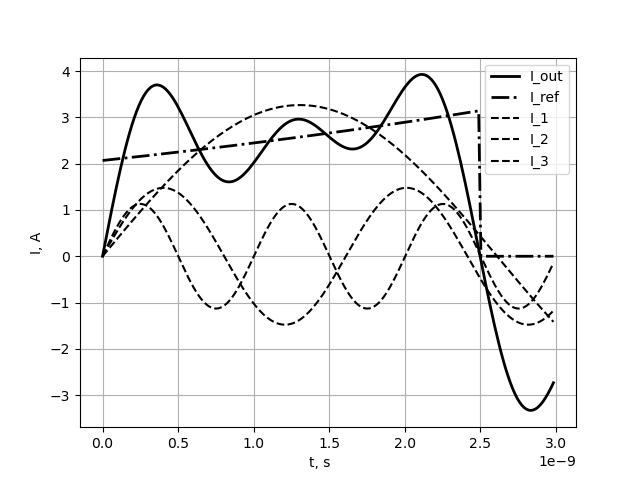}

\caption{Current waveforms of the laser driver with three LC branches. The total output current $I_{out}$ through the laser diode in comparison to the reference optimal trajectory $I_{ref}$. The currents through individual LC branches are shown as well $I_1, I_2, I_3$, $I_{out} = I_1 + I_2 + I_3$}
\label{fig:multi_resonant_LC_waveforms}
\end{figure}

Even though the current waveform present in Figure \ref{fig:multi_resonant_LC_waveforms} is not sufficiently close the exponential one when using a circuit with only three branches, it is obvious that there is an waveform shape improvement in comparison to just one LC branch (i.e. conventional capacitively coupled circuit \cite{gan_appnote}). The turn off current slew rate is much faster as well than comparing with a sinusoidal current.
Additionally, as any capacitively coupled circuit, this driver turns off naturally, so no precise timing is required.

\subsection{Implementation 3: Push-Pull driver and Parallel Capacitor}

A push-pull constant voltage $V$ driver gives a linearly increasing current during the precharge state due to board and package inductive parasitics $L$:

\begin{equation}
I(t) \approx  \frac{V}{L} t
\end{equation}

One simple way to improve the shape of the current is to use a capacitor across the load.

\begin{figure}[H]
\centering
\includegraphics[width=0.6\textwidth]{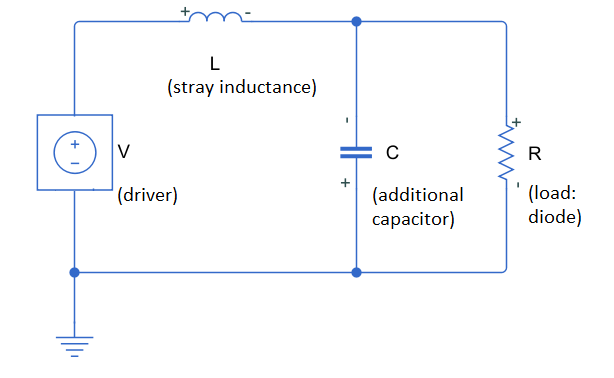}

\caption{Driver topology with capacitor connected across the laser diode}
\label{fig:capacitor_across_concept}
\end{figure}

Let's consider an equivalent circuit presented in Figure \ref{fig:capacitor_across_concept}. The laser diode is modeled as a resistor R. The push-pull driver is a voltage source V. The board and package inductive parasitics are lumped into inductance L.

By calculating a transient response of the circuit, the current through the resistor R in underdamped case ($L < 4 R^2 C$) is given as follows:

\begin{equation}
I(t) = \frac{V}{R} \left ( 1 - e^{-t / \tau} \cos (a t / \tau) - \frac{e^{-t / \tau}}{a} \sin (a t / \tau) \right )
\end{equation}

where

\begin{equation}
\tau = 2 R C, \; a = \sqrt{\frac{4 R^2 C}{L} - 1}
\end{equation}

\begin{figure}[H]
\centering
\includegraphics[width=0.8\textwidth]{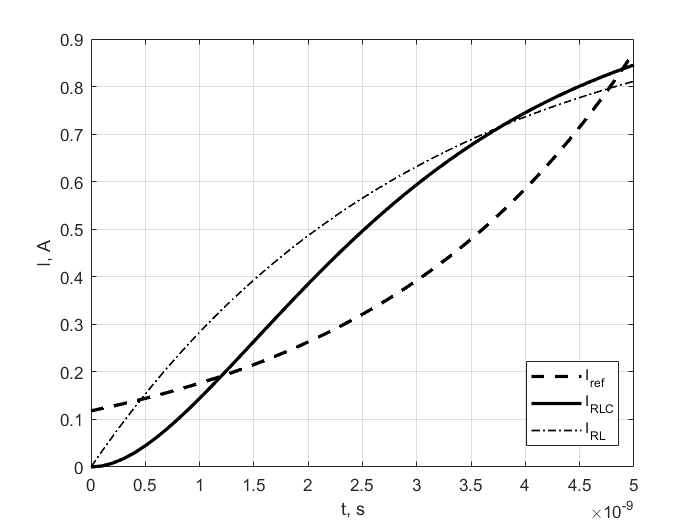}

\caption{The current response of the proposed topology ($I_{RLC}$) and of the circuit with stray inductance $L$ only $I_{RL}$ for load $R = 5$ ohm, $C = 150$ pF and $L = 15$ nH and $V = 5$ V in comparison to optimal trajectory $I_{ref}$ calculated for a 1W 850nm diode for pulse duration of 5 ns}
\label{fig:capacitor_across_waveforms}
\end{figure}

A comparison of the current waveforms with and without capacitor added is given in Figure \ref{fig:capacitor_across_waveforms}.

Although, the current profile obtained in this topology is not exactly exponential, this is considerable improvement achieved by minimal circuit change.

\begin{figure}[H]
\centering
\includegraphics[width=0.8\textwidth]{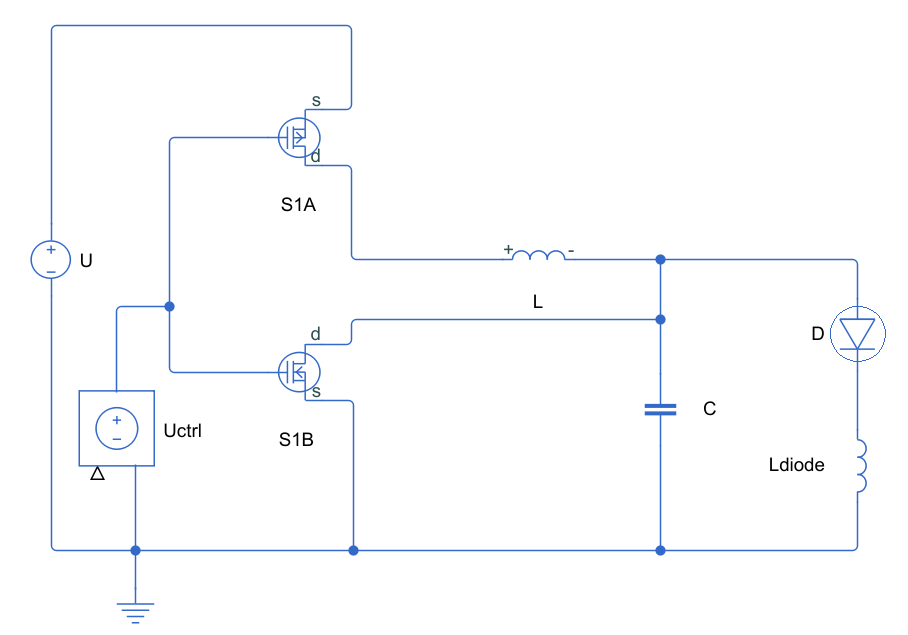}

\caption{Practical circuit of the laser driver of proposed topology}
\label{fig:capacitor_across_circuit}
\end{figure}

Taking the approach practically, it worth noting that in order to achieve a fast turn off current slew rate, the pull down transistor S1B should be connected to the diode by much lower inductance than the pull up transistor S1A, as it shown in the Figure \ref{fig:capacitor_across_circuit}.

In order to achieve the higher inductance in pull up path, this inductance can be artificially increased by a longer PCB trace.

\subsection{Implementation 4: Push-Pull Driver and Saturating Inductor}

When the magnetic flux in the discrete inductor is approaching its saturation limit, the current starts rising very fast.
This effect can be used to generate approximately exponentially increasing current. 

More precisely, the inductance variation of a saturating inductance can be approximated as \cite{ind_saturation}:

\begin{equation}
L(I) = L_{sat} + \frac{L_0 - L_{sat}}{2} \left (1 - \frac{2}{\pi} \arctan (\sigma \cdot (I - I_1) ) \right )
\end{equation}

where $L_{sat}$ is an inductance in fully saturated regime, $L_0$ an inductance far from saturation and $L(I_1) = (L_0 + L_{sat})/2$, $\sigma$ is model parameter.

\begin{figure}[H]
\centering
\includegraphics[width=0.7\textwidth]{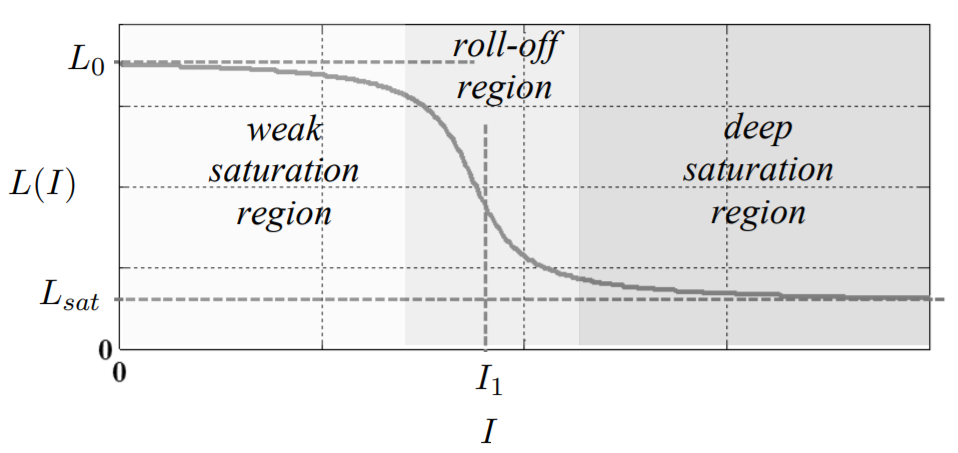}

\caption{Typical inductance vs current curve of an inductor (from \cite{ind_saturation})}
\label{fig:inductance_vs_current}
\end{figure}

\begin{figure}[H]
\centering
\includegraphics[width=0.9\textwidth]{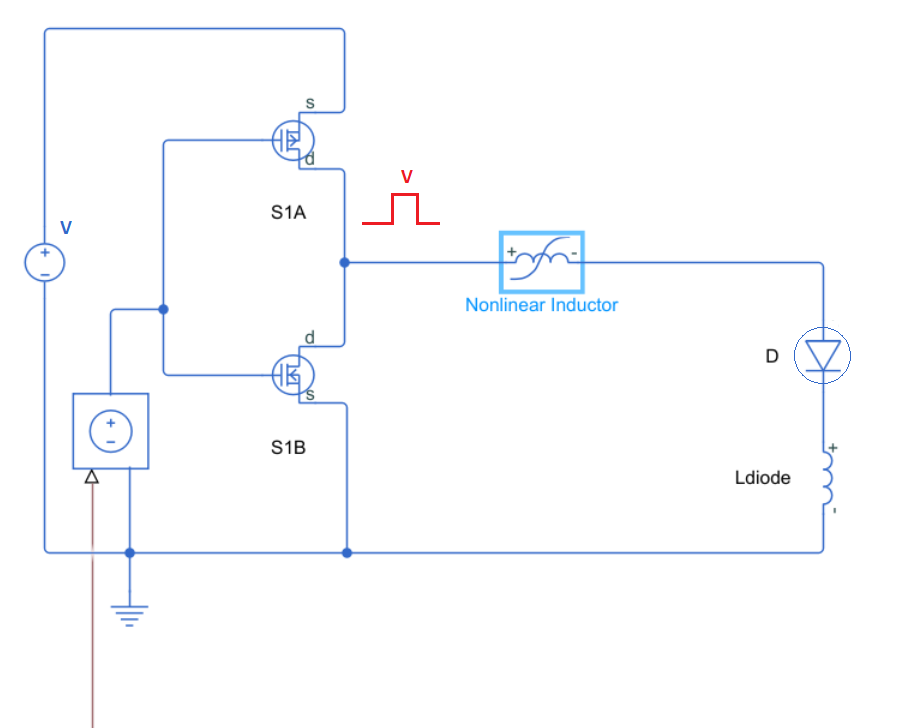}

\caption{Practical laser driver circuit shaping using saturating inductor}
\label{fig:saturating_inductor_circuit}
\end{figure}

The practical circuit implementing this idea is shown in Figure \ref{fig:saturating_inductor_circuit}. The laser driver circuit consists of a half bridge made by transistors S1A and S1B. The output of the bridge is connected to a laser diode D which has intrinsic parasitic inductance $L_{diode}$. The nonlinear saturating inductor in between the driver and the diode provides current waveform shaping. 

The output of the bridge is producing rectangular voltage pulses of amplitude $V$.
Let's assume negligible and constant voltage drop across the laser diode, as well as ideal switches in the bridge. After switching the S1A on, the saturating inductance becomes connected to a constant voltage source $V$, the current $I$ will be governed by the following equation:

\begin{equation}
\frac{dI}{dt} = \frac{V}{L(I) + L_{diode}} 
\end{equation}

\begin{figure}[H]
\centering
\includegraphics[width=0.9\textwidth]{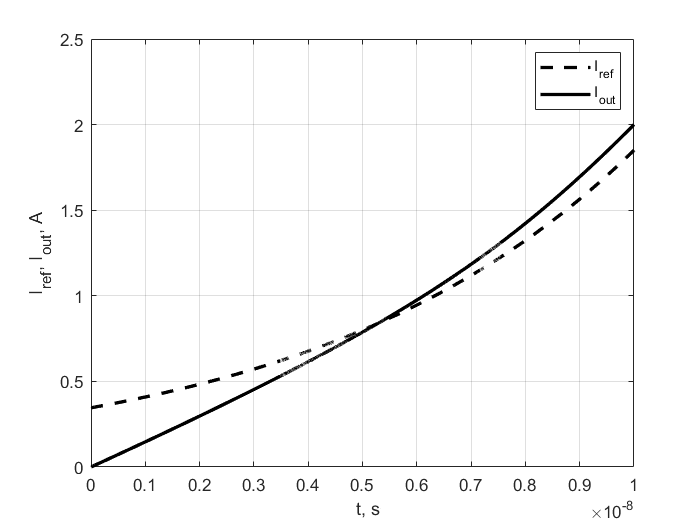}

\caption{Simulation of the laser diode driver circuit with saturating inductor}
\label{fig:saturating_inductor_waveforms}
\end{figure}

The simulated current output of the circuit $I_{out}$ in comparison to the reference optimal current $I_{ref}$ are shown in Figure \ref{fig:saturating_inductor_waveforms}. The pulse duration $T = 10$ ns, the saturating inductor has nominal inductance of $L_0 = 35$ nH, and saturated inductance of $L_{sat} = 5$ nH, the parasitic diode inductance is $L_{diode} = 5$ nH. The optimal trajectory is calculated for a fitted model of the LDX-3820-860, the 860nm wavelength, 200 $\mu$m bar size, 8 W continuous power laser diode in TO-9 package and 750 mA threshold current.

It worth noting that, the saturation current should be pretty low in order to generate the current profile suitable for the laser diodes.
There is a contradiction between required inductance and resulting saturation current, very small inductors have very high saturation currents, because

\begin{equation}
I_{sat} \sim \dfrac{N \cdot B_{sat} \cdot S}{ L }
\end{equation}

where $N$ is the number of turns, $S$ is the inductor cross-section area and $B_{sat}$ is saturating flux density (a material property), L is inductance.

This leaves the method with a limited applicability (especially for small power laser diodes), albeit theoretically possible.

\section{Experimental comparison of the sine-wave-shaped and triangular-shaped current laser drivers}

In this section we compare two practical laser driver circuits used in industry. One generates a sinewave shaped pulse current through the diode and another one uses a direct connection to the push-pull stage driver.
Both circuits were running in the same conditions and operated to achieve the same average power and pulse quality metric $\rho$.

The laser diode Sharp GH0637AA2G with nominal power 700mW and 638nm wavelength is used to benchmark both circuits. This particular laser diode has an isolated package that doesn't connect to either the anode or cathode, which allows benchmarking both of circuits.

\begin{figure}[H]
\centering
\includegraphics[width=0.9\textwidth]{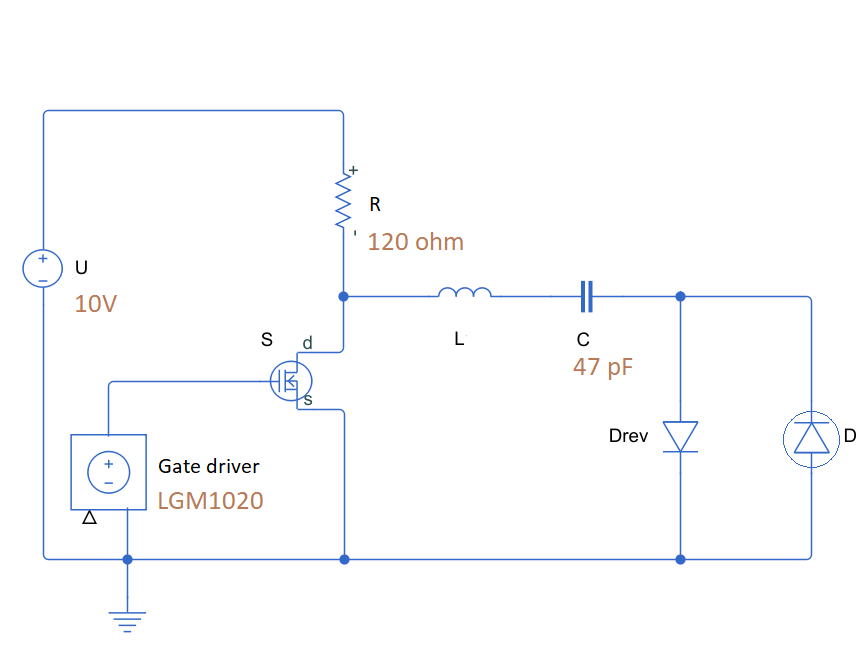}

\caption{Resonant laser driver circuit used for the efficiency comparison}
\label{fig:resonant_laser_driver_circuit}
\end{figure}

\begin{figure}[H]
\centering
\includegraphics[width=0.9\textwidth]{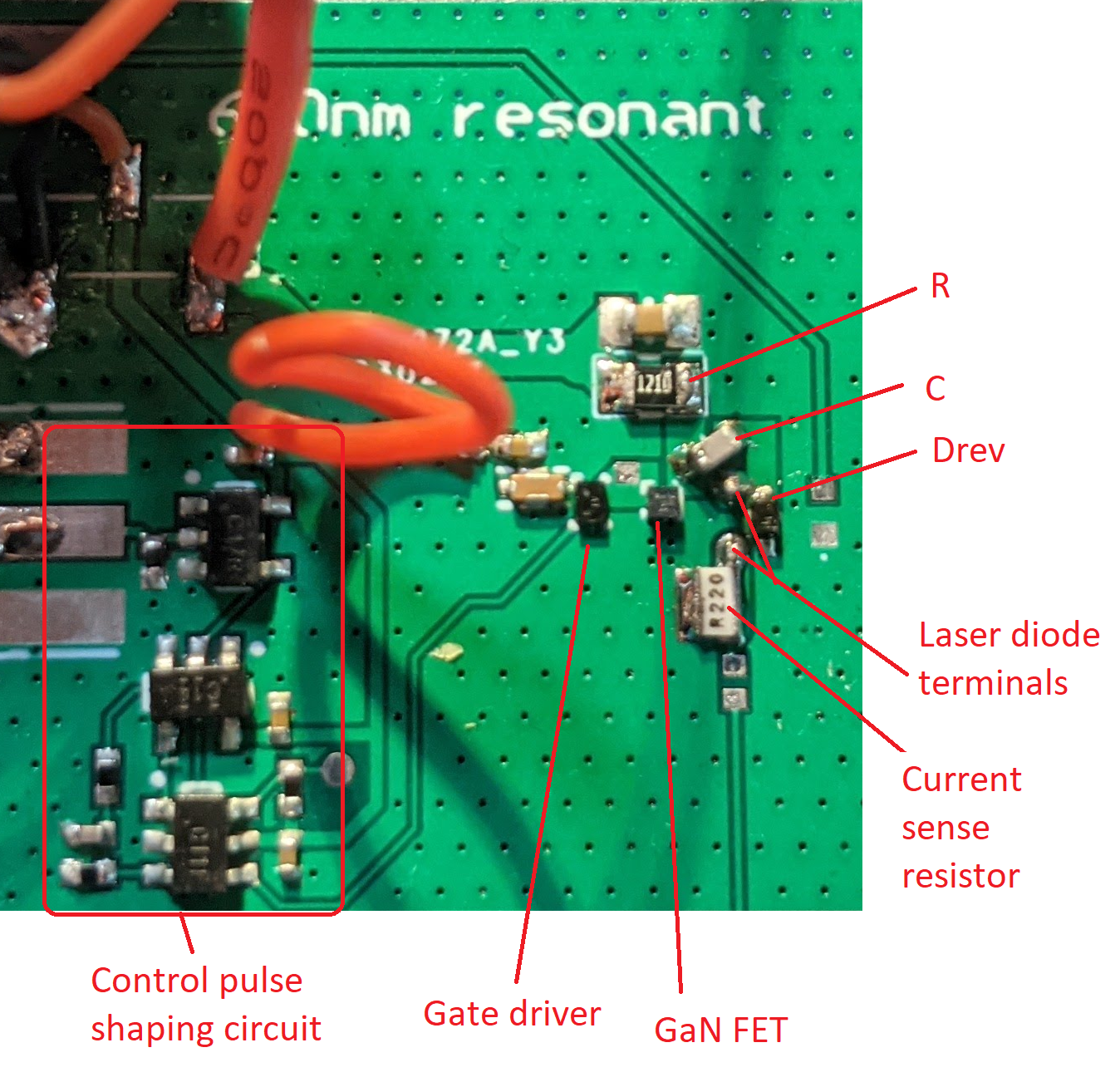}

\caption{Resonant laser driver circuit prototype PCB (the laser diode is on the opposite side of the circuit)}
\end{figure}

The sinewave current driver is based on the very well known resonant capacitive discharge laser driver design \cite{gan_appnote}. The circuit (Figure \ref{fig:resonant_laser_driver_circuit}) operate as follows. The GaN transistor S starts in the off-state and the capacitor C is been charged through R and Drev.
After the capacitor C is fully charged the gate driver turns the transistor S on by external command. Since the S is biderectionally conducting the capacitor C  discharges through the laser D and parasitic inductance L. The C and L form a resonant network, hence the current through the diode D rings sinusoidally. The first half wave of the current ignites the laser diode D in gain-switching mode. During the consecutive ringing the energy stored in the capacitor C dissipates quickly and the next period usually is not enough to ignite the diode, but it contributes to the inefficiency of the circuit. The gate driver turns off the transistor S sometime after the first half wave of the current.

As it seen, the efficiency consideration of this circuit consists of the two power soucres: the gate driver power supply and the main power supply U of the resonant circuit. So the efficiency can be calculated as:

\begin{equation}
\eta = \frac{P_{optical}}{P_{driver} + P_{main}}
\end{equation}

where $P_{optical}$ is output optical power, $P_{driver}$ is electrical power consumed by the gate driver, $P_{main}$ is electrical power consumed by the main resonant circuit from the voltage source U.

In this particular circuit implementation, the LGM1020 gate driver by TI is used to drive the EPC2036 GaN transistor.

\begin{figure}[H]
\centering
\includegraphics[width=0.9\textwidth]{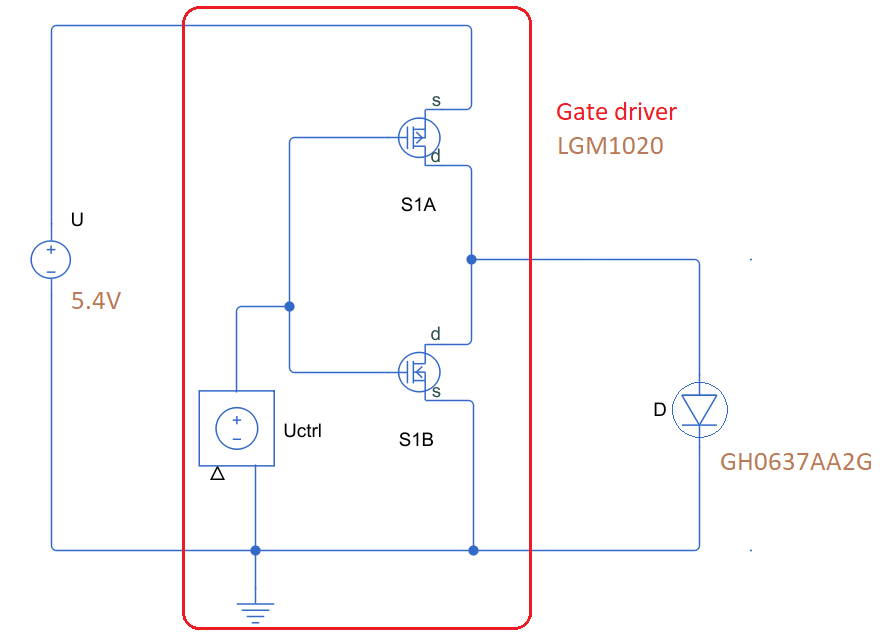}

\caption{Direct push-pull driver circuit used for the efficiency comparison}
\label{fig:direct_laser_driver_circuit}
\end{figure}

The push-pull circuit (Figure \ref{fig:direct_laser_driver_circuit}) is based on the same LGM1020 gate driver, which consists of the high speed push-pull output stage. The laser is simply connected directly to the push-pull output of the driver. The current pulse width and amplitude is regulated both by adjusting pulse width of the control signal and power supply voltage of the gate driver within the acceptable ranges of the LGM1020.

\begin{figure}[H]
\centering
\includegraphics[width=0.9\textwidth]{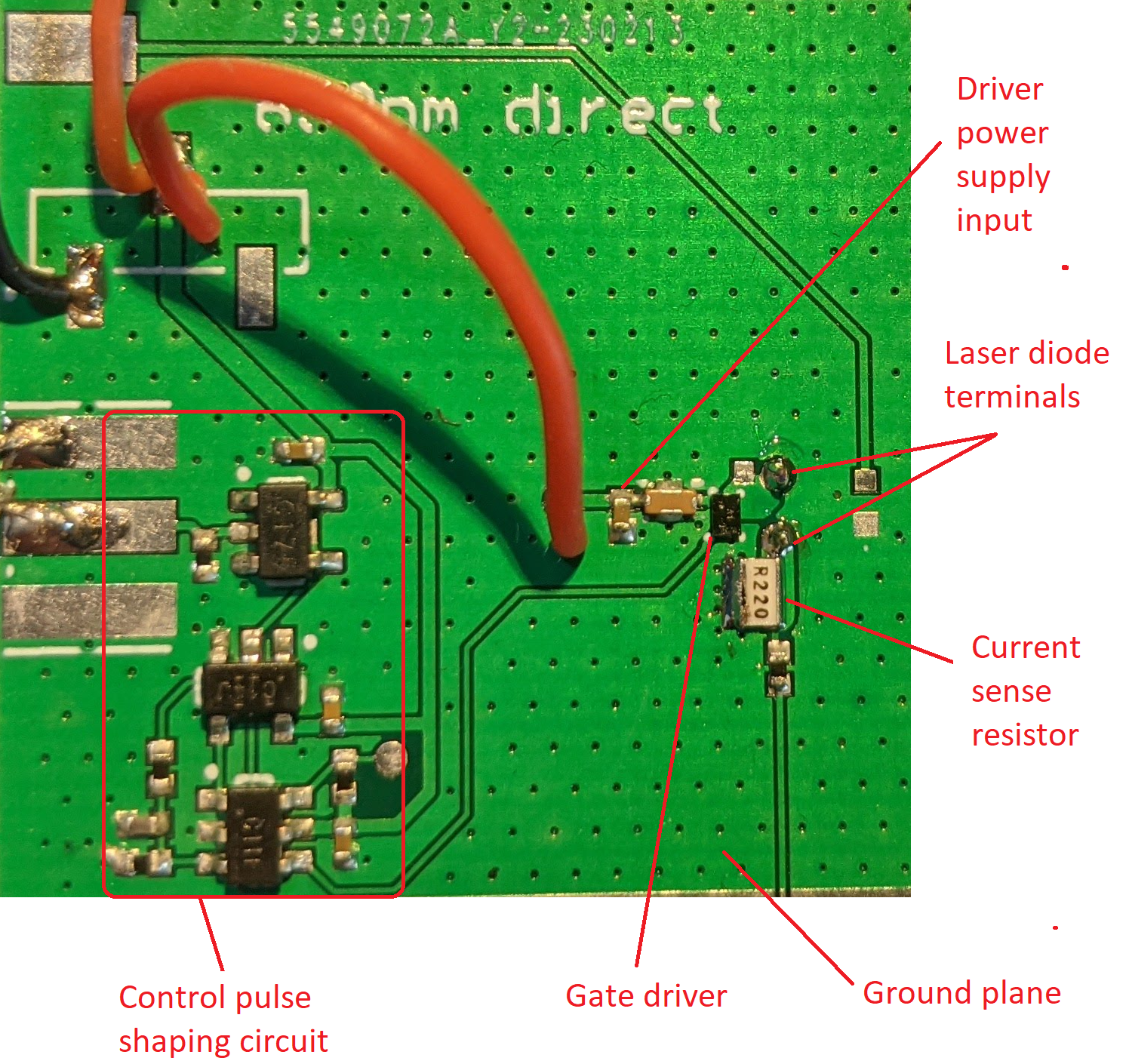}

\caption{Direct push-pull driver circuit prototype PCB (the laser diode is on the opposite side of the circuit)}
\end{figure}

The comparison of the two circuits is summarized in the table \ref{table:comparision}. Both circuits were running at 2 MHz repetition rate and tuned to produce 0.55 mW of average power. In the both cases the power supply voltage of the gate driver was set to 5.4V, the maximum voltage the driver supports. The voltage of the resonant circuit was set to 10V.

\begin{table}[H]
\centering
\begin{tabular}{c|c|c|c|c|c|c}
\hline
Circuit & $P_{driver}$, mW & $P_{main}$, mW & $P_{optical}$, mW & Efficiency, \% & $\rho$, 1/ns & FWHM, ps \\ \hline
Resonant & 19.6 & 35 & 0.55 & 1.0 \% & 6.08 & 110 \\ 
Direct & 16.85 & 0 & 0.55 & 3.3 \% & 6.473 & 120 \\ 
\end{tabular}
\caption{Measured characteristics of the resonant and direct drive circuits.}
\label{table:comparision}
\end{table}

\begin{figure}[H]
\centering
\includegraphics[width=0.9\textwidth]{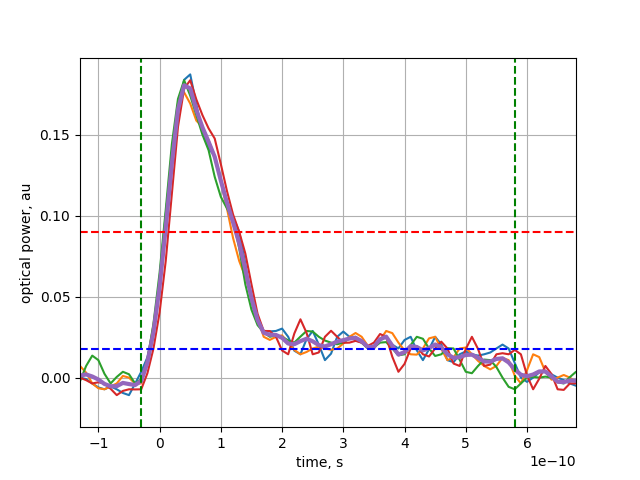}

\caption{Optical pulse waveforms acquired for the resonant circuit}
\label{fig:resonant_circuit_pulses}
\end{figure}

\begin{figure}[H]
\centering
\includegraphics[width=0.9\textwidth]{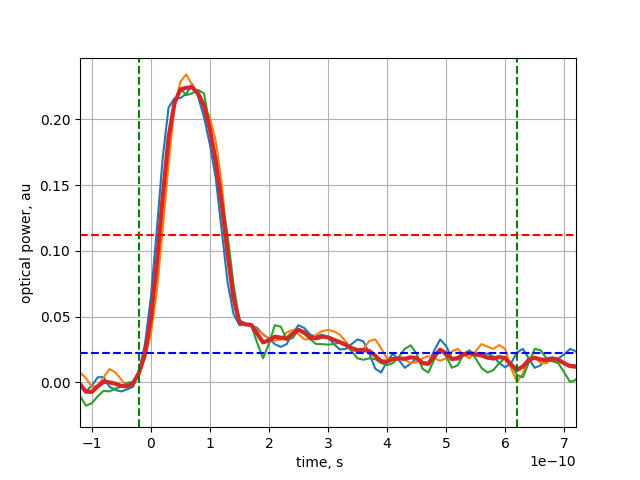}

\caption{Optical pulse waveforms acquired for the direct push-pull circuit}
\label{fig:direct_circuit_pulses}
\end{figure}

The optical pulse is acquired by Tektronix DPO73304SX high speed scope together with the DX12CF ultrafast photodiode sensor by Thorlabs. The acquired and averaged waveforms are presented in the figures \ref{fig:resonant_circuit_pulses} and \ref{fig:direct_circuit_pulses}. The limits of integration used to calculate the $\rho$ is shown by the green vertical lines, and the half-maximum level is shown as red horizontal line.

Evidently the efficiency of the push-pull driving circuit is higher than the sinewave one, while matching the same average power level and calculated $\rho$. 
There are a couple of differences that contribute to that:

- in both circuits the gate driver consumes almost the same power, but the direct driver circuit doesn't have the high voltage part and secondary power supply,

- the direct driver circuit doesn't have dissipative parts as the precharge resistor R,

- the direct driver circuit has more optimal current shape than the resonant one.

\section{Conclusion}

This paper contributes to both the theoretical and practical development of the high-speed semiconductor laser driver circuits for gain-switching operating mode

\begin{itemize}
  \item It is shown that exponentially rising current in the form of $I(t) = A \cdot \exp(t / \tau_N)$ is optimal in terms of resistive ohmic loss for driving the semiconductor laser into the gain-switching mode.
  \item A metric $\rho$ to quantify the quality of laser operation that measures the similarity of a generated optical pulse to the delta function is proposed.
  \item Several circuit implementations to approximate exponentially rising current are developed.
  \item An experimental comparison between state-of-the-art sinewave resonant driver circuit and directly driven laser is performed that favors the latest variant of the driver.
\end{itemize}

The future research work will be directed towards implementation and evaluation of the proposed circuits using mixed-signal chip semiconductor technology.

\section*{Acknowledgment }
Author is grateful to Dr. Han Ban for technical discussions and suggestions during the research and design work, and Mikhail Panin for assembing hardware prototypes of laser drivers.


\begin{thebibliography}{9}

\bibitem{steppulse_modulation} 
Li, B., Sun, C., Ling, Y., Zhou, H. and Qiu, K., 2019. Step-Pulse Modulation of Gain-Switched Semiconductor Pulsed Laser. Applied Sciences, 9(3), p.602.

\bibitem{direct_drive_prechage} 
Arslan, S., Shah, S.A.A. and Kim, H., 2019. Power Efficient Current Driver Based on Negative Boosting for High-Speed Lasers. Electronics, 8(11), p.1309.

\bibitem{direct_drive2}
Zeng, Z., Sun, K., Wang, G., Zhang, T., Kulis, S., Gui, P. and Moreira, P., 2017. A compact low-power driver array for VCSELs in 65-nm CMOS technology. IEEE Transactions on Nuclear Science, 64(6), pp.1599-1604.

\bibitem{resonant_example} 
Huikari, J.M., Avrutin, E.A., Ryvkin, B.S., Nissinen, J.J. and Kostamovaara, J.T., 2015. High-energy picosecond pulse generation by gain switching in asymmetric waveguide structure multiple quantum well lasers. IEEE journal of Selected Topics in Quantum Electronics, 21(6), pp.189-194.

\bibitem{resonant_example2} 
Nissinen, J. and Kostamovaara, J., 2015. A high repetition rate CMOS driver for high-energy sub-ns laser pulse generation in SPAD-based time-of-flight range finding. IEEE Sensors journal, 16(6), pp.1628-1633.

\bibitem{resonant_example3} 
Liero, A., Klehr, A., Hoffmann, T., Prziwarka, T. and Heinrich, W., 2016, October. GaN laser driver switching 30 A in the sub-nanosecond range. In 2016 11th European Microwave Integrated Circuits Conference (EuMIC) (pp. 460-463). IEEE.

\bibitem{resonant_example4} 
Ma, Y.S., Lin, Z.Y., Lin, Y.T., Lee, C.Y., Huang, T.P., Chen, K.H., Lin, Y.H., Lin, S.R. and Tsai, T.Y., 2019, February. 29.6 A Digital-Type GaN Driver with Current-Pulse-Balancer Technique Achieving Sub-Nanosecond Current Pulse Width for High-Resolution and Dynamic Effective Range LiDAR System. In 2019 IEEE International Solid-State Circuits Conference-(ISSCC) (pp. 466-468). IEEE.

\bibitem{optimal_control} 
Naidu, D.S., 2002. Optimal control systems. CRC press.

\bibitem{crest_factor}
Palicot, J. and Louët, Y., 2005, September. Power ratio definitions and analysis in single carrier modulations. In 2005 13th European Signal Processing Conference (pp. 1-4). IEEE.

\bibitem{ind_saturation} 
Di Capua, G. and Femia, N., 2015. A novel method to predict the real operation of ferrite inductors with moderate saturation in switching power supply applications. IEEE Transactions on Power Electronics, 31(3), pp.2456-2464.

\bibitem{optimal_design} 
Rafailov, E.U. and Avrutin, E., 2013. Ultrafast pulse generation by semiconductor lasers. In Semiconductor lasers (pp. 149-217). Woodhead Publishing.

\bibitem{gan_appnote}
Glaser, J.S. (2019). eGaN FETs for Lidar – Getting the Most Out of the EPC9126 Laser Driver.

\bibitem{comparision1}
Ashok, P. and Madhan, M.G., 2017. Optimum electrical pulse characteristics for efficient gain switching in QCL. Optik, 146, pp.51-62.

\bibitem{comparision2}
Ashok, P. and Madhan, M.G., 2019. Performance analysis of various pulse modulation schemes for a FSO link employing gain switched quantum cascade lasers. Optics \& Laser Technology, 111, pp.358-371.

\bibitem{comparision3}
Krishna K, Madhan M, Ashok P. Study of gain switching in vertical cavity surface emitting laser under different electrical pulse inputs. Def. Sci. J. 2020 Sep 1;70(5):538-41.

\bibitem{comparision4}
Ashok, P., 2018. Particle swarm optimization approach to identify optimum electrical pulse characteristics for efficient gain switching in dual wavelength quantum cascade lasers. Optik, 171, pp.786-797.

\end{thebibliography}
\end{document}